\documentclass{article}

\usepackage{arxiv}

\usepackage[utf8]{inputenc} 
\usepackage[T1]{fontenc}    
\usepackage{hyperref}       
\usepackage{url}            
\usepackage{booktabs}       
\usepackage{amsfonts}       
\usepackage{nicefrac}       
\usepackage{microtype}      
\usepackage{cleveref}       
\usepackage{lipsum}         
\usepackage{graphicx}
\usepackage{natbib}
\usepackage{doi}
\usepackage{enumitem}

\title{User-Centered-Design as an Empty Signifier in the Context of Developing Digital Applications}

\newif\ifuniqueAffiliation
\uniqueAffiliationtrue

\ifuniqueAffiliation 
\author{ \href{https://orcid.org/0000-0003-3432-2860w}{\includegraphics[scale=0.06]{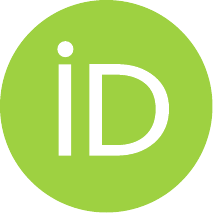}\hspace{1mm}Murat~Sariyar}\thanks{murat.sariyar@bfh.ch} \\
	Department of Engineering and Computer Science\\
	Bern University of Applied Sciences\\
	Bern, 3002, Switzerland \\
	\texttt{murat.sariyat@bfh.ch} \\
}
\else
\usepackage{authblk}

\newbox{\orcid}\sbox{\orcid}{\includegraphics[scale=0.06]{orcid.pdf}} 
\author[1]{%
	\href{https://orcid.org/0000-0000-0000-0000}{\usebox{\orcid}\hspace{1mm}Murat~Sariyar\thanks{\texttt{murat.sariyat@bfh.ch}}}%
}

\affil[1]{Department of Engineering and Computer Science, Bern University of Applied Sciences, Bern, 3002, Switzerland}
\fi

\renewcommand{\shorttitle}{User-Centered-Design as an Empty Signifier}

\hypersetup{
pdftitle={User-Centered-Design as an Empty Signifier},
pdfsubject={q-bio.MS},
pdfauthor={Murat~Sariyar},
pdfkeywords={User-Centered Design, Empty Signifier, Laclau, Lacan, Sociotechnical Systems, Human-Computer Interaction},
}

\begin{document}
\maketitle

\begin{abstract}
To reduce cycles of rejection and redesign -- especially in the absence of clear acceptance criteria and the diversity of possible development paths -- User-Centered Design (UCD) has become a central methodology in computer science, emphasizing the integration of user perspectives throughout the entire system lifecycle. Despite its widespread adoption, however, UCD remains conceptually ambiguous and theoretically underdeveloped. This paper addresses that gap by drawing on the theories of Ernesto Laclau and Jacques Lacan to analyze UCD as a potential empty signifier: a term that gains rhetorical power precisely through its semantic openness. We argue that this ambiguity enables UCD to unify diverse and sometimes conflicting expectations under a shared label, which both empowers participatory design practices and conceals underlying tensions. Acknowledging UCD as an empty signifier allows for a more critical engagement with its practical and symbolic functions, revealing how it can foster inclusivity, empathy, and user empowerment, but also how it risks ideological capture and conceptual dilution. This theoretical reframing opens new pathways for reflection and renewal within sociotechnical system design.
\end{abstract}


\section{Introduction}

The growing dependence on digital systems -- both in professional and everyday life -- places increasing importance on user acceptance. To mitigate costly cycles of user dissatisfaction and redesign, \emph{User-Centered Design} (UCD) has emerged as a key methodological framework within computer science and related fields. UCD promotes the continuous integration of user perspectives from the early stages of project development through deployment. According to \citep{luna2017}, UCD is "a process framework that makes a system usable and understandable by accounting for end-users' needs, wants and constraints, through the whole product cycle". While such definitions are intuitively persuasive and practically useful, they often lack theoretical precision. As a result, UCD is frequently communicated through illustrative examples and case studies rather than through formal conceptual articulation \citep{smith2007}.

The prevalence of illustrative over conceptual definitions reflects a broader issue: UCD's role as a flexible, but loosely bounded, framework. It is often treated less as a clearly delimited methodology and more as a design ethos -- akin to Agile or Lean approaches -- where principles of responsiveness, participation, and iteration are emphasized. While this flexibility facilitates widespread adoption, it also impedes critical reflection. As \citet{twidale2021} note, terms like "usability" and "user needs" are often invoked without consensus on their meaning, yet serve to legitimize a variety of design decisions. The field has responded to this conceptual instability with hybrid frameworks that blend UCD with formal specification and agile workflows, but the rhetorical power of "putting the user first" often outpaces its analytical clarity \citep{fricker2015, lindgaard2006}.

One outcome of this conceptual openness is that UCD is often understood more as a toolkit or development ethos -- similar to Agile methodologies -- than as a theory of interaction or design. Indeed, software engineering increasingly acknowledges that traditional requirements engineering is ill-equipped to capture evolving user needs, which often shift dynamically in response to ongoing development and use \citep{fricker2015}. Consequently, there is growing interest in hybrid approaches that combine formal system specification with UCD's iterative, exploratory methods. Such integration calls for a deeper theoretical understanding of what UCD actually represents -- and why it retains such rhetorical and practical appeal \citep{lindgaard2006}.

Historically, UCD emerged in the 1980s as an evolution of usability engineering, transforming into a structured process involving continuous user feedback and iterative design refinement \citep{norman1986}. Its typical implementation follows recognizable stages: defining the target user group, analyzing user tasks, benchmarking against competitors, conducting design walkthroughs, performing usability evaluations, and assessing outcomes against predefined benchmarks. While none of these steps are unique to UCD, the framework's emphasis on user participation at every stage signals a shift in design ideology -- one that appears to gain legitimacy precisely through its openness.

This prompts a theoretical question: what accounts for UCD's durability and cross-disciplinary appeal, despite its conceptual fuzziness? One possible answer lies in the theory of the empty signifier, introduced by \citet{laclau2005}. Laclau argues that certain terms gain political and discursive power precisely through their ambiguity. As signifiers with no fixed referent, they unify otherwise disparate demands by symbolizing an imagined totality. In earlier work, Laclau demonstrated how such signifiers mobilize collective action by naming perceived injustices or exclusions \citep{laclau1990}.

Similar dynamics can be observed in adjacent discourses. For example, in the field of sustainability, \citet{brown2016} shows how vague but emotionally charged terms like "injustice" serve to articulate failures of dominant paradigms to acknowledge future generations or marginalized voices. These signifiers gain strength by condensing multiple, often conflicting, grievances under a shared rhetorical banner -- thereby opening the possibility for what \citet{badiou2007} describes as a  "true event": a rupture that reconfigures prevailing systems of meaning.

Laclau's theory is deeply influenced by Lacanian psychoanalysis, particularly the notion of the master signifier, which functions to anchor meaning in otherwise unstable symbolic structures \citep{hook2016}. Both draw on Saussure's foundational distinction between signifier and signified, and his insight that meaning is produced through differential relations, not intrinsic content \citep{saussure2011}. For Laclau and Lacan, a master or empty signifier temporarily arrests this fluidity, representing a whole (e.g., "the user") without needing to define its parts.

In this paper, we adopt this theoretical lens to critically examine UCD as a potential empty signifier. As \citet{dopp2019} note, "UCD is a diverse, innovative field that remains highly variable in terms of language and approaches". Similarly, \citet{chen2010} argue that design should not only meet customer needs but also address environmental and social wellbeing -- a framing that stretches the scope of UCD far beyond traditional usability. These perspectives suggest that UCD operates less as a stable methodology and more as a discursive placeholder for a wide array of often incompatible concerns.

The diversity of interpretations regarding what constitutes "user needs" or "user behavior" makes it difficult to construct a unified theory of UCD. This motivates our hypothesis that UCD functions as an empty signifier: widely adopted, variably interpreted, and rhetorically potent across disciplinary boundaries. By treating UCD in this way, we seek to clarify its role in both theory and practice. Such a perspective does not dismiss the value of UCD but rather situates it within a broader discursive economy, allowing us to reflect more critically on its promises, limitations, and future trajectories.

The remainder of this paper is structured as follows: (1) we outline the theoretical criteria for identifying empty signifiers; (2) we apply these criteria to the discourse surrounding UCD; and (3) we propose directions for future research into the conceptual and practical roles of UCD within sociotechnical systems.

\section{Criteria for Categorizing Terms as Empty Signifiers}
\label{sec:two}

\subsection{Lacan's Master Signifier}

\noindent To critically assess UCD's conceptual status and discursive role, we turn to the theories of Lacan and Laclau. Their work on signifiers, subjectivity, and symbolic organization offers powerful tools for understanding why certain terms -- despite their ambiguity -- retain such enduring rhetorical and practical force. Lacan initiated a decisive linguistic turn in psychoanalysis by integrating structuralist and philosophical insights from Freud, Hegel, and Lévi-Strauss \citep{leavy1977}. Central to his theory is the idea that the subject is not a self-transparent entity but is constituted through and by language -- specifically, through signifiers. One of Lacan's most cited definitions holds that a signifier is "that which represents a subject for another signifier" \citep{hyldgaard2009}. This formulation highlights the relational nature of meaning: any articulation of identity or subjectivity emerges only within a chain of signifiers that defer and displace one another. Yet these signifying chains are inherently incomplete. They cannot exhaustively represent the subject, leaving gaps or "structural lacks" that give rise to desire -- the ongoing compulsion to symbolize what resists symbolization \citep{zizek2009}.

Lacan attributes this symbolic insufficiency to a constitutive lack in the so-called \emph{Other} -- the symbolic order as the locus of meaning, law, and social norms. This Other is not merely another person but a structuring principle instantiated in signifiers such as "the Law of the Father" or "God". These figures function as anchoring points in discourse, yet their inability to fully signify what they represent mirrors the subject's own internal lack. The failure to internalize the Other -- its values, expectations, or guarantees -- reveals that the Other itself is incomplete. This insight, though often unconscious, destabilizes the subject's faith in any totalizing symbolic system and propels the ceaseless generation of new meanings. Nonetheless, subjects continue to refine their grasp of experience by deploying signifiers that attempt to carve the real more precisely within the symbolic domain.

Yet if meaning arises only through the differential relations among signifiers, language risks becoming an endless chain without a stable point of reference. To halt this slippage, discourse requires a privileged term that arrests the deferral of meaning. This is the function of the \emph{master signifier} -- a term that grounds the symbolic order not through intrinsic meaning, but through structural centrality and self-reference \citep{hook2016}. Master signifiers like \emph{money}, \emph{freedom}, or \emph{society} appear to carry weight or fullness, yet their semantic content is circular or void. For instance, money represents value but also serves as the means to acquire value, including itself. Freedom, especially in its Kantian formulation as "the will that wills itself", similarly refers back to its own operation. These terms derive authority not from clarity but from their rhetorical centrality within discourse.

The master signifier thus operates as a kind of structural illusion: it appears to fix meaning, yet in doing so it conceals its own emptiness. Importantly, all signifiers within a discourse are affected by this nodal point, which functions as a dislocating force that can disrupt the differential relations between terms \citep{zizek2006}. This destabilizing function is not a flaw but a constitutive feature of language. The master signifier represents the system of signification as a whole precisely because it holds this paradoxical position -- it both organizes and destabilizes, structures and fractures. Lacan often theorized the subject and society in terms of such contradictory signifiers: entities that simultaneously instantiate structure and its failure \citep{zizek2013}. In this respect, the master signifier resembles Kant's notion of a "regulative principle"  -- a heuristic that guides understanding without being directly constitutive of knowledge.

Given that symbolic systems can never fully map onto lived experience, they must remain open to revision and displacement. Lacan foregrounds the linguistic figures of \emph{metonymy} and \emph{metaphor} as mechanisms through which symbolic articulation extends its reach and plasticity \citep{mellard1991}. Metonymy operates by substitution through contiguity (e.g., "in-his-hand" for "in-hostage"), while metaphor enables more radical semantic displacement (e.g., "diapers" standing in for "infancy"). These rhetorical devices do not merely embellish language but serve to redistribute or stretch the semantic field organized by the master signifier. The productivity of these tropes underscores the contingent and open-ended nature of meaning.

This line of thought is later expanded by Ernesto Laclau, who reinterprets the master signifier through the concept of the "empty signifier". For Laclau, such terms acquire political potency precisely because they are semantically indeterminate. Their emptiness allows them to unify disparate demands and identities under a shared banner, enabling hegemonic articulation. In this sense, the master signifier's emptiness is not a deficiency but a precondition for its mobilizing force.

\subsection{Laclau's Empty Signifier and a Methodological Summary}

Building on Lacan's insights into the structural role of the master signifier, Laclau recasts this function in a political and discursive context -- where symbolic indeterminacy enables hegemonic articulation. Laclau's theory of society centers on the interplay between hegemony and the empty signifier, concepts that explain how collective identities are constructed and how political meaning is stabilized \citep{laclau1990}. Hegemony, in this framework, is not merely domination but the contingent fixing of meaning around particular demands, a process that secures temporary unity in a field of difference \citep{butler2000}. Central to this dynamic is the empty signifier -- a term that gains symbolic power not through precise definition, but through its capacity to absorb diverse and even contradictory demands, serving as a placeholder for a universal or totalizing horizon of meaning.

Rather than clarifying or resolving contradictions, the empty signifier functions precisely by suspending them. It organizes chains of equivalence that link disparate grievances or identities under a shared banner. \citet{gross2016} provides an ethnographic example of this in rural Latin American healthcare, where the term Aire is deployed when no other diagnosis can adequately explain an illness. In these cases, Aire allows both patients and practitioners to symbolically contain uncertainty and preserve a sense of control: \begin{quote}

In the situation where every other explanation of illness fails, the possibility to diagnose the condition as 'Aire' enables the patient to symbolically maintain control over the ailment. [...] It serves as a general metaphor for physiological and especially social and psychological disequilibrium for the demarcation of which people lack more specific vocabulary.
\end{quote} 

What is striking here is that the semantic incoherence of Aire does not undermine its value. On the contrary, its ambiguity enables it to perform a vital symbolic function -- sustaining a sense of order, continuity, and meaning in the face of diagnostic failure. This is emblematic of how empty signifiers work: they resolve nothing conceptually but enable symbolic endurance and alignment.

Crucially, empty signifiers do not signal objectivity or stability. Instead, they reveal the antagonistic structuring of discourse, the fault lines along which collective identities are formed. As Laclau writes, "the social never manages to fully constitute itself as an objective order" \citep{laclau1990}. Social cohesion is animated not by consensus or clarity but by investment -- libidinal, imaginative, and affective -- into symbols that appear to unify what is in fact contingent and fragmented \citep{hook2016}. For example, political signifiers like liberal democracy may come to stand against totalitarianism, condensing entire chains of equivalence: inequality = liberty versus equality = totalitarianism \citep{laclau2001}. Such alignments simplify complex terrains of conflict while masking their constructed nature.

Over time, these chains of equivalence may become metaphorically stabilized -- fixed in meaning, elevated in abstraction, and stripped of visible contingency. Contestation then involves reactivating metonymy -- undoing the metaphor by reintroducing partial, contextual, or associative links that expose its constructed character \citep{payne2016}. As Laclau notes: \begin{quote}
 The dissolution of a hegemonic formation involves the reactivation of that contingency: the return from a 'sublime' metaphoric fixation to a humble metonymic association \citep{laclau2014}.
\end{quote}

This dialectic between metaphorization and re-metonymization will be central to our later discussion of discursive and practical formations. We aim to contrast the logic of empty signifiers with practices that resist abstraction and instead operate through metonymic specificity, partial connection, and contextual grounding. To assess whether a term functions as an empty signifier, we propose the following methodological criteria, adapted from Lacan and Laclau:

\begin{itemize}
\item Absence of a stable definition, coupled with a general lack of reflection on that absence (Lacan and Laclau)
\item Transition from metonymy to metaphor, in which associative uses become ideologically fixed (Lacan)
\item Antagonistic structuring, where the term helps organize a political frontier or binary (Laclau)
\item Formation of chains of equivalence, where diverse demands coalesce around the term (Laclau)
\end{itemize}

These criteria allow us to identify when a signifier functions hegemonically and to trace how meaning becomes fixed or contested within discursive formations.

\section{Application of the Empty Signifier to the UCD Term}

After applying our four criteria for identifying a term as an empty signifier to the concept of User-Centered Design (UCD), this section explores the implications of viewing UCD through this lens. First, we assess to what extent UCD lacks a standard definition and whether there is a noticeable absence of critical reflection on this fact. Second, we examine the semantic shift in the use of UCD from metonymic to metaphorical. Third, we discuss whether UCD has become or is likely to become politicized in terms of antagonisms. Fourth, we analyze the impact of long equivalence chains built around positive examples on UCD practice. Finally, we outline key theoretical and practical implications that follow from this analysis.

\subsection{No Standard Definition and Lack of Reflection Thereof}

The literature reveals a wide range of definitions for UCD. For example, \citet{luna2017} describe UCD as a "process framework that makes a system usable and understandable by accounting for end-users' needs, wants, and constraints throughout the entire product cycle", based on ISO 9241-210. \citet{dopp2019} define UCD as a "design approach that grounds the characteristics of an innovation in information about the individuals who use that innovation, with a goal of maximizing usability in context". Couture et al. emphasize heuristics such as "match between system and the real world, consistency and standards, flexibility and efficiency of use, and aesthetics and minimal design " \citep{couture2018}. Meanwhile, \citet{graham2019} describe a six-phase UCD process aligned with typical software engineering steps: investigate, ideate, prototype, evaluate, refine and develop, and validate.

Two interconnected conclusions emerge. First, it is difficult to identify what is genuinely novel about UCD when compared to existing processes in software engineering, such as agile methods, which already emphasize user feedback and iterative development. Tools like the System Usability Scale (SUS) are widely used across contexts and are not specific to UCD. Second, the absence of a standard definition allows for broad and sometimes contradictory interpretations, enabling researchers to append additional goals or features without a clear conceptual boundary.

Why is there so little reflection on this conceptual vagueness? Three interrelated epistemological factors can be identified. (i) UCD lacks a cohesive theoretical foundation and is often applied pragmatically across domains to justify practices rather than develop conceptual clarity. (ii) UCD is typically situated within transdisciplinary initiatives that prioritize practical extension over theoretical coherence. In such settings, ad hoc definitions are common, and conceptual precision is not a priority. As \citet{putnam1975} noted, laypeople -- and by extension, interdisciplinary practitioners -- tend to rely on stereotypes rather than scientifically rigorous characterizations. (iii) The foundational ambiguity of disciplines such as computer science affects subfields like UCD. For example, \citet{vallverdu2010} points out that unlike natural sciences, computer science deals more with values than objective truths. This foundational instability hinders the development of paradigms that could anchor reflective discourse and contributes to a permissive "anything-goes" mindset.

\subsection{From Metonymic to Metaphorical Usage}

Despite the lack of a standard definition, UCD is widely used to describe user-centered methodologies. This usage has evolved from denoting specific practices (e.g., usability testing, interaction design) to representing a broader ethos or ideal. Common metonymies for UCD include terms like user-perspective, user-friendly, intuitive, and design thinking. These suggest a user-centered orientation but often lack consistency in how users are actually integrated into design processes.

This broadening of meaning indicates a shift from metonymic to metaphorical use. UCD becomes a symbolic placeholder for all activities framed as user-inclusive, regardless of their depth or rigor. It acts as a metaphor for responsible or systematic design, streamlining communication but reducing conceptual precision. As with other metaphors, frequent use can lead to cognitive economization and a decline in critical engagement. The term's vagueness is not merely a deficiency but a linguistic resource that enables cross-contextual applicability and superficial consensus.

This is especially visible in contemporary labels such as "intuitive AI", "human-centered machine learning", or "ethical tech", which borrow legitimacy from UCD while repackaging it within broader ideological narratives. These slogans often invoke user alignment or moral responsibility without specifying mechanisms of participation or criteria of inclusion. As such, they exemplify how UCD functions metaphorically -- not as a concrete method, but as a diffuse signifier for desirable technological futures. This further amplifies its symbolic reach while deepening its conceptual ambiguity.

\subsection{Politicization Through Antagonisms}

The metaphorical abstraction of UCD opens it up to politicization. By functioning as a floating signifier, UCD allows practitioners to frame practices as either aligned with or deviating from its principles. This paves the way for antagonistic debates over what constitutes legitimate UCD. Following Laclau, such antagonisms emerge when multiple actors offer competing self-descriptions of a practice, leading to struggles over boundaries and definitions.

Although overt antagonisms around UCD are still emerging, critical voices are growing, especially those linking UCD to broader ethical and societal concerns. This politicization is neither inherently negative nor avoidable. It reflects the normative dimensions embedded in design practices and highlights the need for shared frameworks to address value-laden disputes. For instance, the question "What constitutes user-friendly design?" cannot be answered solely empirically but requires ethical deliberation. As \citet{correia2012} argues, argumentation must account for its ethical dimension to achieve rational discourse. Hence, theoretical engagement with antagonisms can help articulate conflicting perspectives, fostering a deeper and more inclusive understanding of UCD.

\subsection{Equivalence Chains of Positive Examples}

In the absence of precise definitions, UCD is often illustrated through chains of positive examples. These include successful case studies, best practices, and heuristics that are presumed to reflect UCD principles. While useful for practical orientation, such examples operate through associative rather than definitional logic. They serve more as exemplars than as conceptual anchors.

This reliance on exemplars parallels psychological models of object categorization \citep{murphy2002}, where understanding is guided by resemblance rather than essence. Accordingly, UCD is not defined by a fixed set of criteria but by a constellation of perceived best practices. As \citet{floyd1989} noted, software systems are not just technical constructs but sociotechnical artifacts embedded in human and machine contexts. This framing positions UCD as a field of evolving experiences, which, while flexible and adaptable, may frustrate attempts at formalization. The tension between flexibility and formalism remains unresolved and must be navigated pragmatically. In line with von Foerster's dictum, "Only those questions that are in principle undecidable, we can decide" \citep{von_foerster2003}, the balance cannot be formally prescribed.

\subsection{Theoretical and Practical Implications}

What follows from these observations for the theory and practice of UCD? Theoretically, it becomes both increasingly difficult and urgent to establish criteria for distinguishing legitimate from illegitimate uses of UCD. As the chain of positive examples grows, so does the risk of conceptual dilution. Yet this very instability can catalyze productive debate, as seen in dialectical models like that of Hegel \citep{wisser1988}. Theories of empty signifiers, such as Laclau's, offer tools to analyze this conceptual flux and guide strategic research questions: Which scientific paradigms apply? Should UCD be integrated into the social sciences? What disciplinary realignments are necessary?

Practically, three conclusions emerge. First, UCD can be invoked either extensionally (to describe practices focused on users) or intensionally (as a stereotype of user-centeredness). In the former, practices like employing SUS scores are labeled as UCD-aligned simply because they involve user feedback. In the latter, UCD becomes a floating reference for values projected onto a design process. This dual usage enables wide applicability but also breeds ambiguity.

Second, metaphorical usage facilitates interdisciplinary collaboration by fostering shared language across contexts. It enables teams to align on general goals -- "putting the user at the center" -- without committing to specific methodologies. However, this flexibility comes at a cost: any step toward concreteness risks exposing value-laden disagreements. Such tensions often surface late in development, when decisions about product features become inescapable. The perceived centrality of the user can then serve as a rhetorical basis for critique, even if the original expectations were never clearly defined.

Third, as UCD continues to operate as an empty signifier, its entanglement with broader ethical and political discourses will deepen. Concepts like sustainability, justice, and freedom are increasingly attached to UCD in research and practice. This creates space for ideological alignment and contestation. UCD alone lacks the rhetorical power to mobilize social critique, but in combination with broader signifiers, it can serve as a vessel for articulating systemic concerns. Consequently, emerging debates will likely frame UCD approaches as "progressive" or "regressive," depending on their alignment with prevailing ethical narratives. As seen in recent works like \citep{twidale2021}, the discourse around UCD is beginning to shift from technical methodology to political positioning.

\section{UCD Applied as an Empty Signifier: Operational Shifts Under Reflexive Acceptance}
\label{sec:applied_empty_signifier}

What changes when all participants in a design process -- developers, researchers, funders, users -- recognize that User-Centered Design (UCD) functions as an empty signifier? That is, what happens when the ambiguity and rhetorical fluidity of UCD are not treated as problems to be resolved, but as intrinsic and even enabling properties of the term? This section explores the practical and epistemic consequences of reflexively accepting UCD's status as an empty signifier. We identify potential advantages and limitations of this stance, focusing on how it transforms coordination, communication, and accountability in real-world settings. Three examples from healthcare, civic technology, and AI ethics help clarify what is gained and what is risked when UCD is embraced in this way.

\subsection{Coordination Without Coherence: Pragmatic Advantages}

When UCD is openly treated as an empty signifier, its main benefit lies in coordination. The term can function as a boundary object, flexible enough to accommodate different interpretations, yet robust enough to support joint action. This facilitates interdisciplinary and cross-sectoral collaboration, especially in complex sociotechnical projects where shared vocabulary is rare.

\paragraph{Example 1: Reflexive UCD in Digital Health.} A public-private consortium develops a chronic care management app involving hospitals, insurers, software engineers, and patient advocates. Rather than attempting to unify all actors under a strict definition of UCD, the consortium acknowledges its semantic openness. Each group articulates what UCD means for them -- data privacy for insurers, intuitive interfaces for engineers, equitable access for advocates -- while maintaining a shared symbolic alignment. This mitigates prolonged definitional debates and enables the project to move forward. The process benefits from a common reference point without demanding conceptual consensus.

This pragmatic pluralism allows UCD to serve as what Luhmann might call a contingency formula a stabilizing semantic tool that absorbs contradiction while preserving communicative functionality \citep{luhmann1998}. As long as actors understand that the term's meaning will differ across perspectives, its emptiness can support rather than hinder joint innovation.

\subsection{Symbolic Efficiency vs. Methodological Ambiguity}

While coordination improves, accountability structures may weaken. If UCD is acknowledged as a floating signifier, then it may lose its capacity to demarcate quality standards or signal design rigor. Without procedural expectations, the invocation of UCD risks functioning as a form of design laundering, a symbolic gesture that legitimizes decisions without operational depth.

\paragraph{Example 2: Civic Tech and the Risk of Performative Inclusion.} In a local government project building a citizen feedback platform, the design team embraces UCD as a shared ideal but makes no effort to implement concrete participatory mechanisms. They justify this on the grounds that all stakeholders agree UCD is a symbol of democratic engagement, not a specific method. While this reflexive stance prevents conflict and streamlines development, it also masks the absence of actual user involvement. Citizens are named as central actors but remain institutionally excluded. The result is a platform optimized for managerial reporting, not for civic dialogue.

Thus, accepting UCD as empty enables efficient alignment at the cost of depth. The term becomes an interface between actors, not between designers and users. This weakens traditional accountability mechanisms such as usability testing, persona development, or field validation, since no group assumes responsibility for defining or enforcing methodological standards.

\subsection{Ethical Framing and the Expansion of Design Discourse}

On a more positive note, accepting the emptiness of UCD can enable ethical expansion. Freed from fixed procedural forms, UCD can evolve into a platform for articulating normative commitments that transcend usability, such as fairness, sustainability, or accessibility. Rather than tying "user needs" to interface preferences, design teams may leverage the symbolic reach of UCD to engage broader ethical imaginaries.

\paragraph{Example 3: Human-Centered AI as Ethical Placeholder.} A research lab developing an NLP model adopts a reflexive approach to UCD. They acknowledge that "user" is an abstraction, but choose to use UCD as a rhetorical anchor for fairness and value alignment. The design team runs deliberative workshops with ethicists, domain experts, and civil society groups, not with end-users per se. Rather than producing classic usability metrics, they generate a value matrix that informs model tuning and risk documentation. UCD here becomes a stand-in for human-centered governance, not empirical design.

By recognizing UCD's semantic openness, the team can adapt its scope to address systemic concerns. The downside is a loss of empirical traction: without concrete user tasks or performance benchmarks, success is harder to evaluate. Nonetheless, the empty signifier enables ethical exploration that more rigid design processes might preclude.

\subsection{Strategic Use vs. Instrumental Application}

Reflexively treating UCD as an empty signifier allows actors to decide when to emphasize flexibility and when to invoke procedural specificity. This strategic use creates room for tailoring design processes to context while still referencing a shared ethos. However, such elasticity risks strategic ambiguity, where terms are invoked for symbolic cohesion but evacuated of operational meaning.

If no actor insists on grounding UCD in specific actions, its invocation may defer or mask decision-making. In this light, the distinction between acting under a signifier and acting through a method becomes essential. Design teams must ask not only what UCD signifies, but how it shapes practice in a given context.

\subsection{Is Non-Reflexive UCD Preferable?}

One might ask whether UCD works better when its ambiguity goes unacknowledged -- when it silently coordinates teams without requiring critical reflection. There is something to be said for tacit consensus: if teams proceed smoothly under the shared illusion that they all mean the same thing by UCD, they may avoid paralyzing debate. This is akin to Wittgenstein's language games \citep{wittgenstein2009} agreement in practice need not require agreement in definitions.

Yet such unreflected use may conceal systemic biases or power asymmetries. Reflexive acceptance at least surfaces the contested nature of design decisions and encourages epistemic humility. Whether that leads to better outcomes depends on context. In low-risk environments, pragmatic ambiguity may suffice. In high-stakes domains,surfacing the symbolic structure of UCD may be a prerequisite for ethical accountability.

\subsection{Summary: Emptiness as a Design Resource}

Accepting UCD as an empty signifier alters its role from prescriptive method to symbolic infrastructure. This reframing changes how actors coordinate, how they justify actions, and how they distribute responsibility. It yields advantages in flexibility, alignment, and ethical scope, but risks weakening empirical grounding and accountability.

Whether this is a gain or a loss depends on whether design systems are judged by their conceptual clarity or their capacity to hold together diverse actors and agendas. The paradox is that UCD may do more work when it is recognized as symbolically empty -- so long as its emptiness is strategically and reflexively managed, not naively presumed or cynically exploited.

\section{Discussion}

Reframing User-Centered Design (UCD) through the concept of the empty signifier offers a novel perspective on its rhetorical flexibility, ideological power, and practical limitations. One pathway for expanding this reflection is to situate UCD within broader paradigms -- such as sustainability, social justice, or inclusivity -- where user needs intersect with collective concerns. Several works already hint at this trajectory \citep{brown2016, sellers2013, wever2008}. These studies underscore that user interaction cannot be separated from environmental and societal systems, since both fundamentally shape human experience. UCD, in this view, becomes not just a design principle but a mediating force between individual behavior and systemic impact. As Wever et al.\ (2008) observe, ``the way users interact with a product may strongly influence the environmental impact of a product.'' If UCD is extended to consider such externalities, it could serve as a vehicle for addressing climate and ethical challenges from within the design process itself.

Beyond sustainability, the analysis of UCD can be meaningfully extended to other foundational notions in computing, such as knowledge, information, and intelligence. Here, theoretical insights from structuralist and post-structuralist thinkers, including Saussure, Lacan, and Laclau, offer valuable tools. Saussure's concept of the \emph{sentiment de la langue} \citep{swiggers2016} suggests that linguistic understanding is not reducible to explicit knowledge but involves implicit, socially mediated intuitions. Applying this to UCD highlights how knowledge about users is always partial, constructed, and framed by linguistic structures. Definitions of knowledge -- whether as "justified true belief" or as "the use of data to solve how-questions" -- are themselves subject to epistemic instability. This aligns with the idea that knowledge may also function as an empty or master signifier, especially when tethered to contested notions such as truth. The enduring influence of Gettier's critique \citep{gettier1963} underscores that even highly formalized definitions can fail to displace deeply ingrained ambiguities.

These observations open a bridge to the social sciences, where the role of language in shaping technical systems is increasingly recognized. Concepts such as Luhmann's systems theory \citep{qvortrup1996} and Deleuze and Guattari's notion of assemblages \citep{deleuze1987} provide additional frameworks for understanding UCD not as a closed method, but as a discursive node within a broader sociotechnical assemblage. Assemblages blend machinic and enunciative components -- hardware, protocols, roles, expectations, metaphors -- and thus help reveal how UCD functions both as a design tool and as a symbolic form \citep{kleinherenbrink2020}. This dual role invites contributions from both insiders (designers, engineers) and critical outsiders (sociologists, philosophers), fostering the kind of transdisciplinary dialogue that can support meaningful interventions.

Interestingly, applying the theory of empty signifiers to UCD also feeds back into theoretical refinement. For instance,  \citet{glynos2007} list five conditions for the emergence of an empty signifier: availability, credibility, articulation by actors, unequal power, and historical embeddedness. Yet these conditions are not sufficient. Some central terms -- such as object-oriented programming -- fulfill all five, yet retain clear operational definitions and structured implementations. This suggests that new criteria are needed to distinguish between conceptually "loaded" but stable terms, and truly empty signifiers whose power lies in their openness. Without this nuance, the risk is high that the theory becomes a tool of indiscriminate critique rather than a framework for productive differentiation.

\section{Conclusion}

This paper has applied the theories of Lacan and Laclau to deepen the conceptual understanding of User-Centered Design (UCD) as a discursive formation within computer science and sociotechnical design. We identified four core characteristics that mark UCD as an empty signifier: (i) the absence of a stable definition and lack of reflection on that ambiguity; (ii) the shift from metonymic reference to metaphorical abstraction; (iii) the emergence of antagonistic interpretations and value-laden debates; and (iv) the construction of long equivalence chains comprising loosely related best practices and examples.

From this perspective, UCD's strength lies in its ability to unify heterogeneous design goals and stakeholder expectations. Its weakness, however, lies in the very same openness that permits overextension, conceptual dilution, and strategic ambiguity. Practically, we identified three implications. First, the dual function of UCD -- as both a technical descriptor and a normative ideal -- facilitates alignment in early-stage projects but often obscures the need for critical decisions later in the process. Second, metaphorical invocations of UCD can defer conflict, leading to latent tensions that resurface at moments of delivery or evaluation. Third, as UCD becomes embedded in wider ideological narratives (e.g., "ethical tech" or "responsible AI"), its use may come to mark political positioning, with some approaches labeled "progressive" and others "exploitative."

Recognizing UCD as an empty signifier does not imply rejection, but reorientation. It invites designers and theorists alike to acknowledge the symbolic power of the term while remaining attentive to its operational consequences. More broadly, it suggests that technical fields benefit from ongoing theoretical reflection -- not to undermine practice, but to enhance its capacity for self-awareness, ethical responsiveness, and critical innovation.

\bibliographystyle{unsrtnat}







\end{document}